# Subtractive Color Mixture Computation[*]

Scott Allen Burns
University of Illinois at Urbana-Champaign (scottb@illinois.edu)

**Overview**

I present an algorithm for computationally mixing screen colors (RGB colors) subtractively, written for a general audience. The question it addresses is, "Given two colors specified by their RGB triplets, what RGB triplet should be used to represent the color that would arise if the two colors were mixed like paint colors, i.e., mixed subtractively?" The only way I can think of doing this in a rigorous way is to employ the math behind how a stimulus (a continuous spectral power distribution) enters our eyes and is transformed into a three-dimensional color sensation by our brain. Once this process has been adequately modeled, subtractive color mixture follows directly. The approach I present here is to convert the RGB colors to spectral reflectance curves, mix the curves using the weighted geometric mean, and then convert the result back to RGB.

**A Disclaimer**

The algorithm described here provides a representative model for subtractive color mixture. The way actual paints mix, for example, is highly dependent upon the particular pigments being used, as well as many other factors. Be aware that you can mix a blue and a yellow paint to get green in one case, and then mix another blue (that appears *identical* to the first blue and has the same RGB value) and another yellow (appearing *identical* to the first yellow, with the same RGB value) and get brown or some other color as a result! There is no way to differentiate between these two different outcomes based on the RGB values of the source colors. Consequently, the colors calculated by my algorithm are plausible, generic, representative color mixture outcomes, and certainly are not predictive of any one specific paint chemistry. Some of the factors that contribute to the unpredictability of paint mixture, not represented by the RGB description of the source colors, include:

o  The varying degree to which light is reflected directly off the top layer of grains of pigment and mix additively in our head.
o  The "glazing" effect that comes from the ordering of layers of transparent paints, giving different results according to the order of application.
o  The shift in hue that results from mixing a color with white paint due to the "undertones" of pigments.
o  The difference in how transparent paints mix compared to opaque paints, due to the additional scattering possible within transparent paint layers.
o  The fact that two visually identical pigments can have very different spectral reflective properties, strongly impacting how they mix with other colors.

---

- The difference in pigment grain sizes, which affect how they respond to tinting and shading actions.

With that said, I believe the algorithm I describe here will give very plausible results in computer graphics applications intended to mimic realistic paint mixture.

**Introduction to RGB Color**

The colors we see on our TVs, computers, and phones, if we look closely enough, are produced by an array of tiny red, green, and blue dots. By varying the relative brightness of these three colors, we can produce a wide range of other colors through "additive" color mixture. We describe these colors using three numbers, the brightness of the R, G, and B dots, typically with whole numbers in the range 0-255, such as (R,G,B) = (217, 15, 145). This is how a program like Photoshop describes its RGB colors. In other settings, the RGB values are treated as floating point numbers the range 0.0 to 1.0, like (RGB) = (0.7263, 0.0112, 0.8956). Programs like Matlab use this convention.

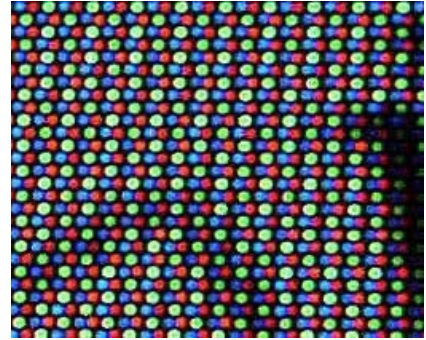

A magnified view of a typical computer monitor. (By ErnstA [CC BY-SA 3.0], via Wikimedia Commons)

The RGB "color space" is a three-dimensional space spanned by the R, G, and B axes. A point within this space represents a unique color. The origin, RGB = (0, 0, 0), represents black (or the darkest color the screen can produce, which is typically quite far from true black). The color RGB = (1.0, 1.0, 1.0) represents the color produced when R, G, and B are at their brightest, and is usually called white. There are many different RGB color spaces, differing in what specific red, green, and blue lights are used for the primaries, and what specific color it considers to be white, called the "reference white." This document will focus on one specific space called the sRGB color space, which is commonly used with RGB display devices. It has a very specific definition[1] that aims to make color images appear consistent across a wide variety of computer screen types.

**Additive vs Subtractive Mixture**

When we overlap the beams of two colored flashlights on a white wall, they mix additively. When we mix two paints of different colors, they mix subtractively. The rules for color mixing are very different in the two cases. Mixing blue and yellow paint, for example, usually gives some sort of green. But mixing blue and yellow light will typically give white or a neutral gray. Mixing red and green paint usually gives a muddy color, but with light, the combination is yellow.

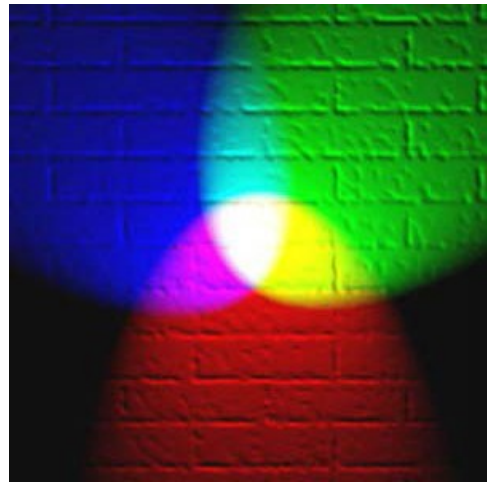

Figure 1 Additive Color Mixture
(By en:User:Bb3cxv [CC BY-SA 3.0] via Wikimedia Commons)

There are computer graphics applications where a programmer needs to mimic subtractive color mixture. An obvious example is a program that teaches how to mix paints, such as [this one](#)[2]. But other applications, like painting programs and photo-realistic scene generation can benefit from being able to model subtractive color mixture. Yet, finding an algorithm for realistically mixing colors this way is surprisingly difficult.

**Why is this a Problem?**

Here are the RGB values for some basic colors:

- red=(1,0,0)
- green=(0,1,0)
- blue=(0,0,1)
- cyan=(0,1,1)
- magenta=(1,0,1)
- yellow=(1,1,0)
- black=(0,0,0)
- white=(1,1,1)

Cyan, magenta, and yellow are often used as subtractive "primaries." Note how well the math works when we *multiply* the RGB values together when mixing them:

- **cyan=(0,1,1)** times **magenta=(1,0,1)** is **blue=(0,0,1)**
- **magenta=(1,0,1)** times **yellow=(1,1,0)** is **red=(1,0,0)**
- **cyan=(0,1,1)** times **yellow=(1,1,0)** is **green=(0,1,0)**

This is just what we would expect from a subtractive mix of these colors. Unfortunately, the multiplicative mixture model breaks down pretty quickly for other pairs of colors. Mixing red and yellow, for example, just gives red as the multiplicative result, not orange as we would expect.

**red=(1,0,0)** times **yellow=(1,1,0)** is **red=(1,0,0) ????**

Even worse, it appears that mixing white with any other color has no effect at all! There is no way to produce a tint of a color in this mixing model.

Others have suggested converting the RGB to other color spaces, such as L*a*b* color space (a "perceptually uniform" color space), CMYK color space, or HSV color space, before doing the mixing to help with the subtractive mix computation. In my experience, I'm not aware of any successful attempts at doing this.

The only way I can think of doing this in a robust way is to first understand how the brain sees colors and then adapt this process to the subtractive mixture of RGB-based colors.

**Mathematics of Human Color Perception**

Our eyes are sensitive to a range of electromagnetic radiation oscillating faster than radio waves and microwaves, but slower than x-rays and gamma rays. The wavelength range of visible light is roughly 400 to 700 nanometers (nm) going from the violet end to the red end of the spectrum.

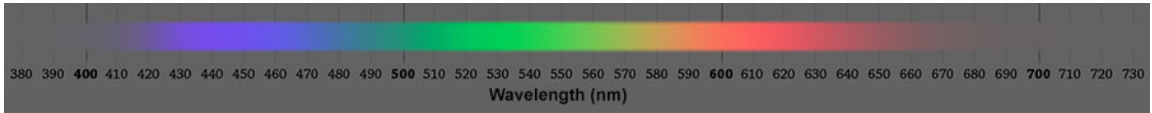

The Visible Spectrum  (By Spigget [CC BY-SA 3.0] via Wikimedia Commons)

Generally, we can describe how an object appears to have a color as a four step process.

1. The object is illuminated by a light source.
2. The object selectively reflects light according to its reflectance properties.
3. The reflected light, or stimulus, passes from the object and into our eyes.
4. The stimulus is processed by our brain to give the impression of a colored object.

Let's look at each step in more detail:

**Light Source**

A **light source** has various levels of power at each frequency that can be summarized in a plot of relative power vs frequency, for example, as shown in the following figure.

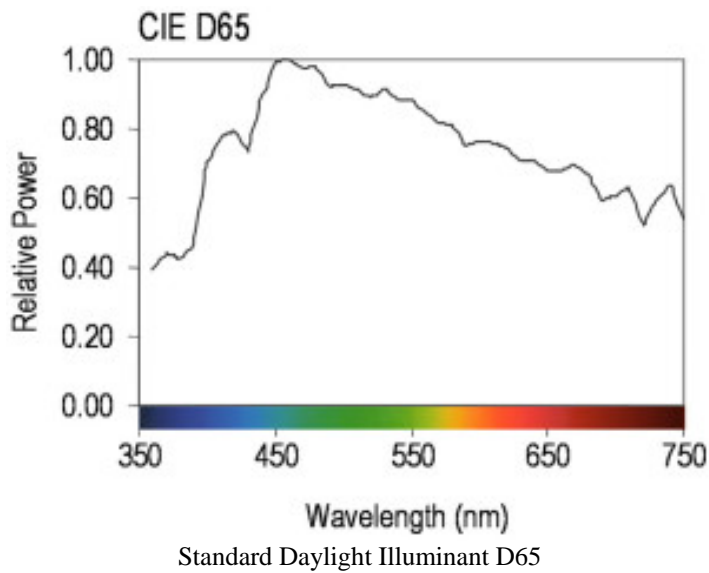

Standard Daylight Illuminant D65

This is one of the "standard" illuminants used in colorimetric studies, called D65. It resembles daylight and is the reference illuminant for the sRGB color space (more on that later). Notice how it peaks more toward the blue end of the spectrum, making it a "cool-er" light source than indoor incandescent illumination, which has much more power on the red end of the spectrum.

**Reflectance Curve**

The reflectance properties of an object are conveniently summarized in a **reflectance curve**, which describes the fraction of incoming light that is reflected at each wavelength. This can be measured using a spectrophotometer. For example, the reflectance curve shown in the following figure was measured from a bright red object:

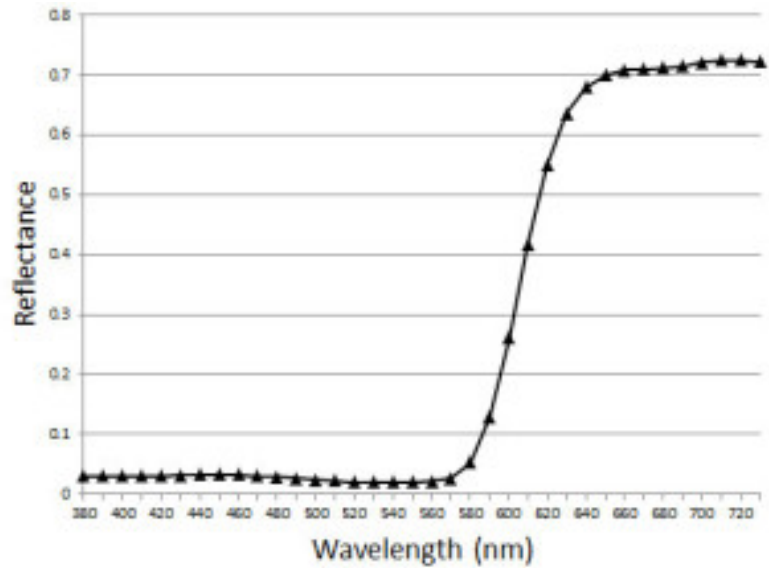

Reflectance Curve for Munsell Red Sample (5R 4/14)

Notice how this red sample reflects around 70% of the longer wavelengths, and very little of the shorter wavelengths.

Since reflectance is a fraction of reflected light, it falls in the range 0 to 1 at each wavelength. The only exception to this is with fluorescent objects, which can reflect more that 100% of incoming light at certain wavelengths, and therefore, have some reflectance values greater than 1.

**Stimulus**

To get the distribution of light that enters our eyes (the **stimulus**), we simply multiply the illuminant by the reflectance, as shown by the red curve in the following figure.

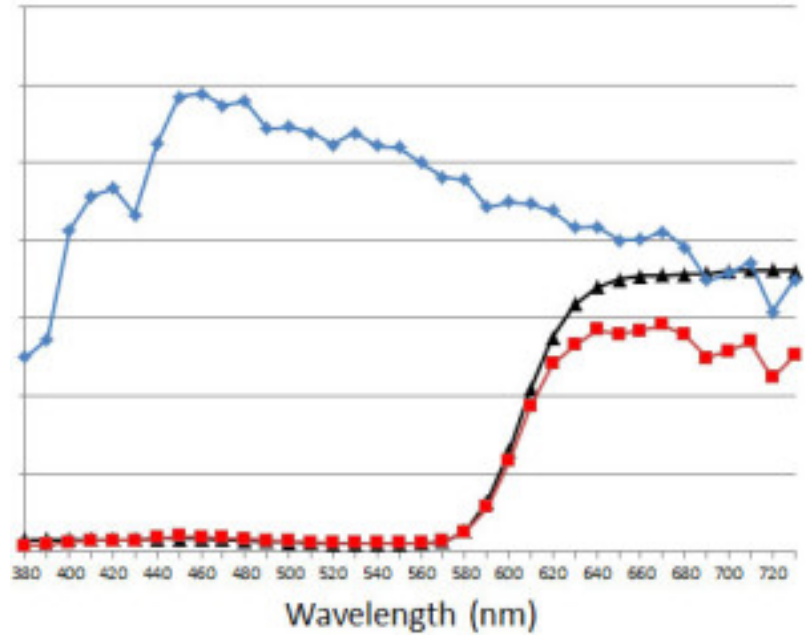

The final light stimulus (red) is the product of the illuminant (blue) and the reflectance (black).

**Color Sensation**

The last step is to describe mathematically how our brain interprets a color stimulus, providing us with a **color sensation**. It can be described by a fairly simple matrix equation:

$$\{rgb\} = [M][A']\{N\}/w, \tag{1}$$

where, $N$ is the stimulus vector with $n$ values, $A'$ is a 3x$n$ "color matching functions" matrix, $M$ is a 3×3 conversion matrix between "tristimulus values" and "linear RGB" values, $w$ is a normalizing factor that is related to the illuminant, and $rgb$ is the 3×1 RGB description of the color before gamma correction, or the "linear RGB" values.

> **More Details…**
> The three rows of the $A'$ matrix comprise three "color matching functions," describing the trichromatic nature of human color vision. They were determined experimentally in several studies over the past century. One in particular, published in 1931, has become the standard model of human color vision.[3] Here is a text file of the $A'$ matrix. It summarizes how a human observer would mix three specified primary colored lights to visually match a fourth target color. When $A'$ is multiplied by the stimulus vector, $N$, and divided by a normalizing factor, $w$, a three-valued color specification is produced, in this case, a set of "tristimulus values," $X, Y$, and $Z$.

The normalizing factor, $w$, is simply the scalar product of the illuminant vector and the second row of $A'$. This makes the tristimulus $Y$ value equal to 1 when the stimulus vector equals the illuminant. The handy thing about normalizing in this way is that the particular units of the stimulus vector no longer matter because they are cancelled out by the normalizing factor.

The tristimulus values, $X$, $Y$, and $Z$ are converted to linear rgb values through a linear transformation matrix, $M$. There are many RGB color spaces, and each has its own RGB "primaries" and reference illuminant (or "white point"). In the case of the sRGB color space (which is referenced to illuminant D65), the $M$ matrix has the values shown here.

Note that I'm using the symbol $A'$ here; the standard $A$ matrix is $n$x3 instead of 3x$n$, and so $A'$ indicates that it has been transposed.

Equation 1 can be further simplified for the special case of obtaining linear RGB values, $rgb$, directly from a reflectance curve, $\rho$. The $M$, $A'$, and $w$ terms can all be combined together ahead of time into a single matrix, $T$. Furthermore, since $N$ is the product of the illuminant and the reflectance curve, $\rho$, the illuminant can be extracted from $N$ and also be combined into the $T$ matrix, leaving only the reflectance curve behind:

$$\{rgb\} = [T]\{\rho\},$$

where $\rho$ is a vector of reflectance values in the range 0 to 1.

> **More Details...**
> The stimulus, $N$, can be expressed as the product of a diagonal illuminant matrix, $\mathrm{diag}(W)$, and a reflectance vector, $\rho$. Putting the illuminant on the diagonal of a matrix will yield a stimulus vector of the proper shape when multiplied by the reflectance vector, $\{N\} = [W]\{\rho\}$. Now, when the stimulus is expressed this way, it is evident that $T = [M][A'][W]/w$.

The size of $T$ and the length of $\rho$ depend on how finely we measure the reflectance curve. Typically, it is measured every 10 nm in the range 380 nm to 730 nm, making $T$ a 3×36 matrix and $\rho$ a 36×1 vector. Here are the specific values for the $T$ matrix (again, for the special case of D65-referenced linear RGB calculation from a reflectance curve).

> **More Details...**
> If you wish to experiment with this matrix in Excel, you can copy the tab-delimited values obtained from the link above, and paste them into Excel (use Paste Special -> Text). Then just matrix-multiply it by the reflectance vector (in Excel: `=MMULT(T_array,rho_vector)` and press ctrl-shift enter).

The final step to get true sRGB values from the linear RGB values is to apply a type of **gamma correction**, also known as "companding" or applying the "color component transfer function." This process will be familiar to anyone using the Levels function in Photoshop, when moving the middle Levels control. The sRGB companding action compresses higher values of RGB and expands lower values, giving a more pleasing and realistic-looking sRGB color space.

> **More Details…**
> A simple, approximate way to do this is to raise each of the $r$, $g$, and $b$ values in $\{rgb\}$ to the 1/2.2 power. The more complicated, actual specifications for sRGB require this transformation: for each $r$, $g$, and $b$ component of $\{rgb\}$, let's generically call it $v$, if $v \leq 0.0031308$, use $12.92v$, otherwise use $1.055v^{1/2.4} - 0.055$. This can be done in Excel with the conditional `=IF(v<0.0031308, 12.92*v, 1.055*v^(1/2.4)-0.055)`. The result of this will be true sRGB values in the range 0 to 1. Multiply them by 255 and round to the nearest integer to get the alternate range of 0 to 255, as used in Photoshop for example (in Excel: `=ROUND(255 * IF(v<0.0031308, 12.92*v, 1.055*v^(1/2.4)-0.055), 0)`). More information on this conversion process can be found at Bruce Lindbloom's highly informative website.[4]

Thus, we now have a way to compute an sRGB triplet given a particular reflectance curve, $\rho$. For the purposes of subtractive color mixture of two sRGB colors, if we know the reflectance curves corresponding to the two sRGB colors, we can perform the subtractive mix on the reflectance curves to arrive at another reflectance curve, and then compute the sRGB values from the mixture reflectance using the equation above. How is the subtractive mix of reflectances achieved? This is discussed next.

**Subtractive Mixture of Two Reflectance Curves**

Suppose we have the reflectance curves for two colors we wish to mix subtractively. After researching the web for insight into this, I found that multiplying the reflectance curves together is a step in the right direction, but will mimic a special type of subtractive mix that results from passing white light through two colored filters, assuming the filters have *transmittance* curves matching the reflectance curves of two colored objects.

But I'm more interested in modeling how paints mix instead of how colored filters act. In the case of two colored objects, I found a website by Bruce MacEvoy called handprint.com that suggests that the "geometric mean" of the two reflectance curves will give a reasonable prediction for the mixture of two watercolor paints in equal proportion.[5] The geometric mean of two reflectances would be computed by multiplying them together and then taking the square root of each term. I decided to investigate further this model of subtractive mixture.

One thing the handprint web site did not address is how to mix colors in unequal proportions, say, five parts of one color to two parts of another. Here, the "weighted geometric mean"[6] is useful. (This idea arose from my previous work in Geometric Programming and the Monospace Method.[7-9]) If we define the weights, $\delta$, to be the fraction of the total that each color contributes to the mix, then the weighted geometric mean (WGM) is

$$\rho_{mix} = \rho_1^{\delta_1}\, \rho_2^{\delta_2}.$$

For example, if we want five parts of the first color and two parts of the second, we would use $\delta_1 = 5/7$ and $\delta_2 = 2/7$. We would then multiply the two reflectance curves together after raising each value in them to the $\delta$ power. Can you see how the case of equal parts of two colors simplifies to the square root of the product of the reflectances?

This extends easily to a mix of more than two colors. Each reflectance curve gets raised to the fraction that it contributes to the whole. A 4:5:6 mix of three colors would be

$$\rho_{mix} = \rho_1^{4/15}\, \rho_2^{1/3}\, \rho_3^{2/5}.$$

As an example, suppose we have two reflectance curves representing a red paint and a blue paint. The graph below shows the mix reflectance coming from a 2:1 mix of red and blue, $\rho_r^{2/3}\, \rho_b^{1/3}$:

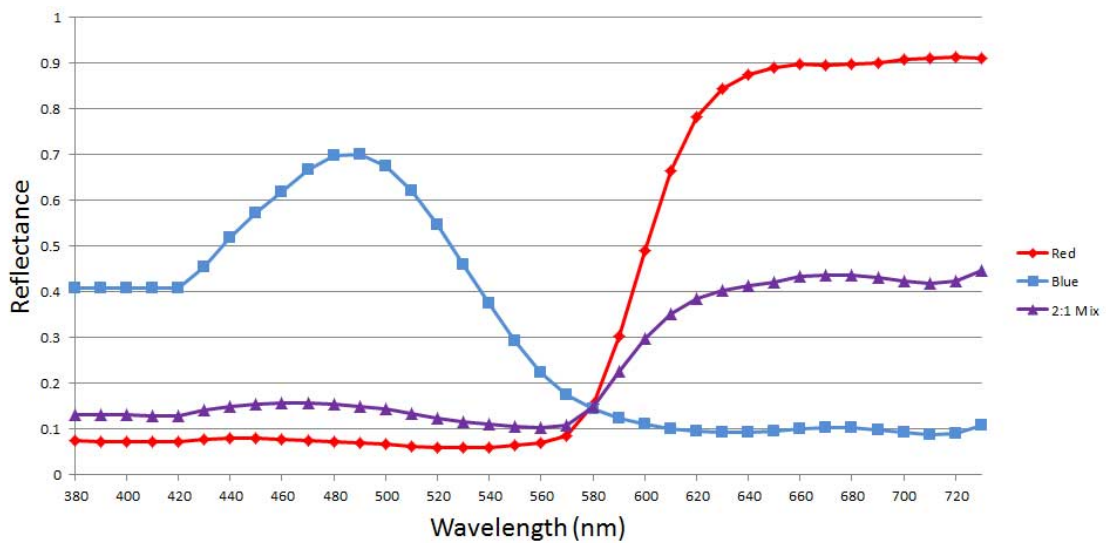

2:1 mixture of red and blue by the weighted geometric mean.

I spent some time examining the characteristics of the WGM mixing method. I found a software program by Zsolt Kovacs-Vajna called rs2color that has a database of reflectance curves of various commercial paints.[10] For example, here are the reflectance curves for 19 of the Liquitex heavy-body acrylic paints[11]:

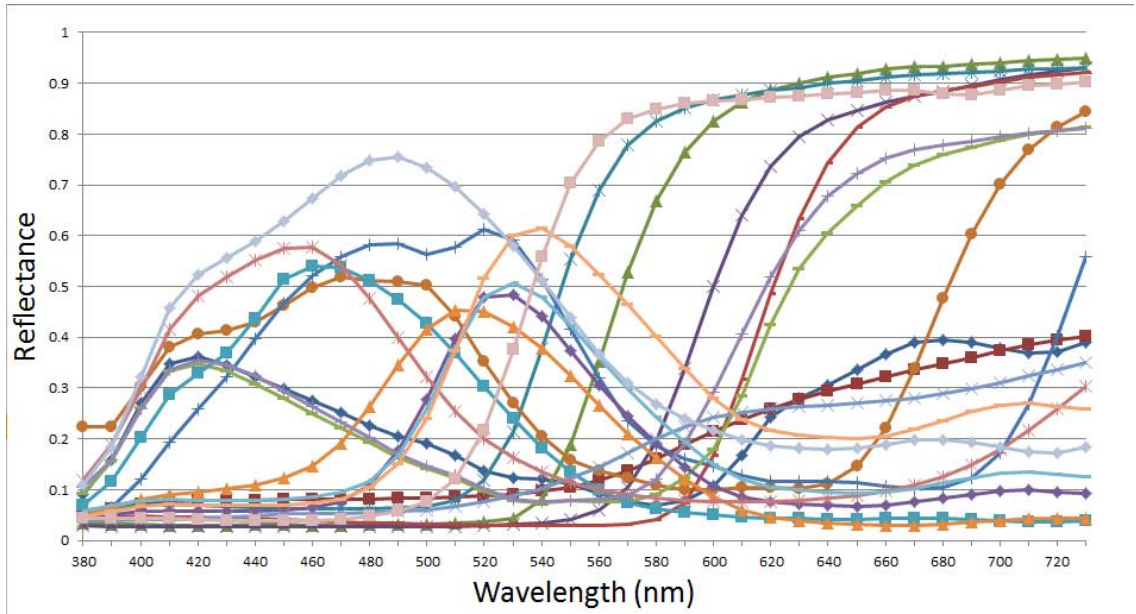
Reflectance curves for 19 Liquitex HB acrylic paints.

Here are the 19 Liquitex paints plotted on a hue/chroma plane in the rs2color program:

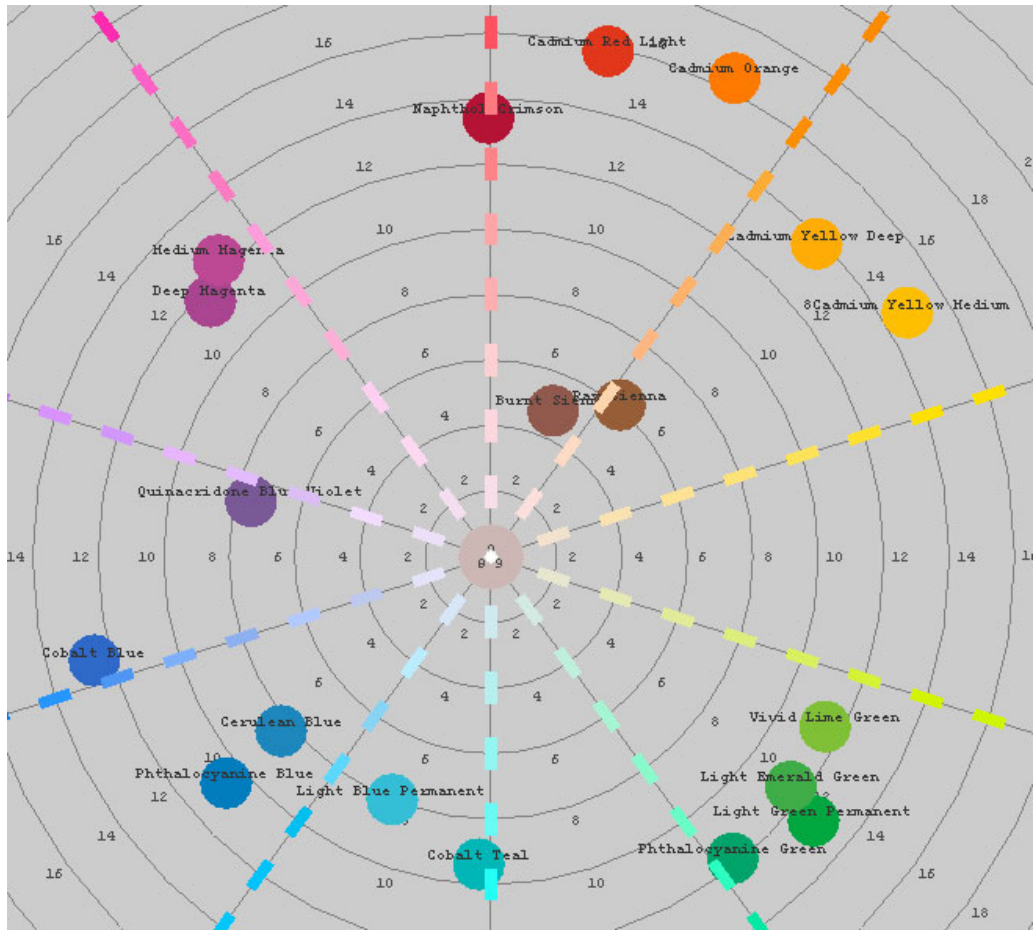

[David Briggs](#) examined the mixing of these 19 Liquitex paints, [plotting the results](#) in rs2color.[12] He describes how colors that are close to cyan, magenta, and yellow mix well with other colors, giving bright, saturated mixtures. He plots "mixing paths" that show the progression of mixing proportions from 1:0 to 0:1, and describes these paths as an "[extroverted octopus](#)" shape, in contrast to the mixing paths he gets when mixing with red, blue, and green, which tend to muddy the mixtures and form an "introverted octopus" shape.

I tried the same thing with the WGM mixing method, generating intermediate mixes 9:1, 8:2, 7:3, ... , 1:9. The following plots show the mixing curves for six of the Liquitex colors, when mixing them with all the others

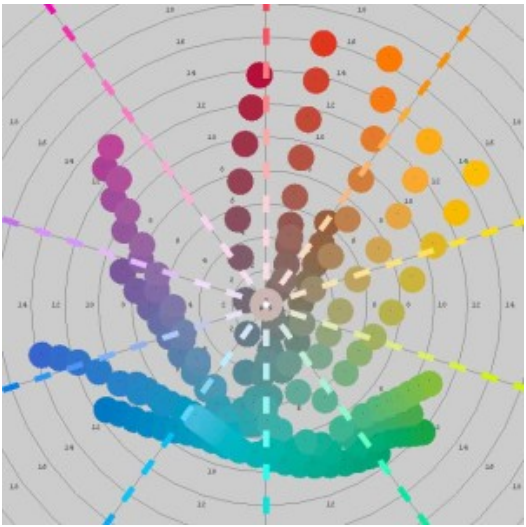

Cyan (#172 Cobalt Teal) mixing with others.

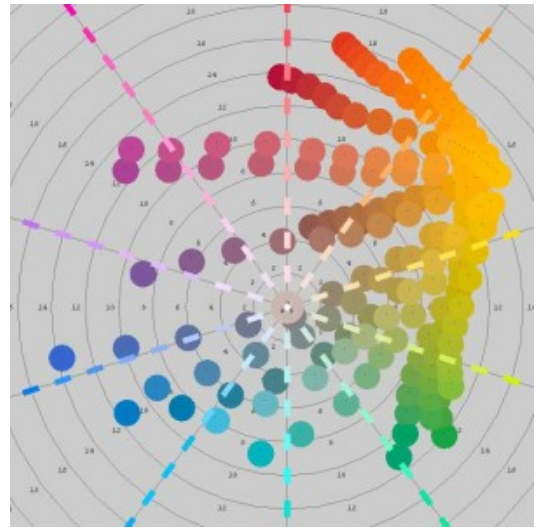

Yellow (#830 Cadmium Yellow Medium Hue) mixing with others.

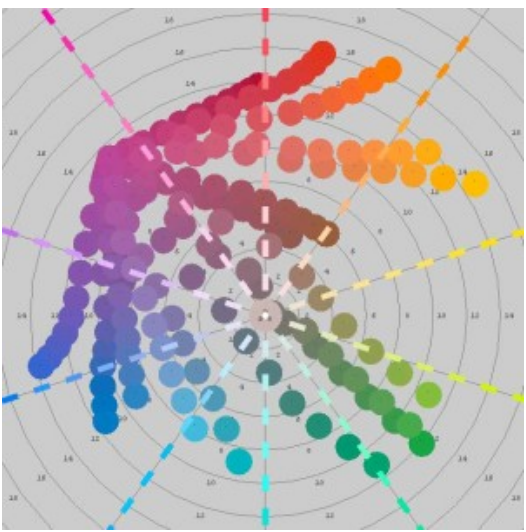

Magenta (#500 Medium Magenta) mixing with others.

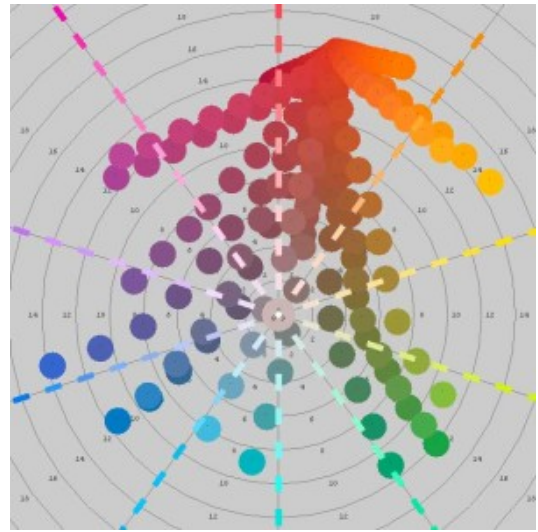

Red (#152 Cadmium Red Light) mixing with others.

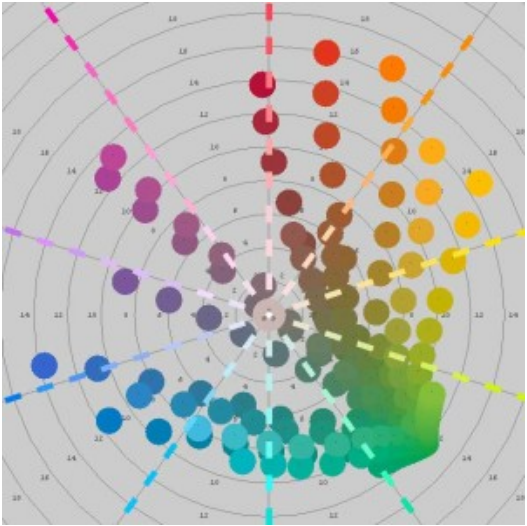 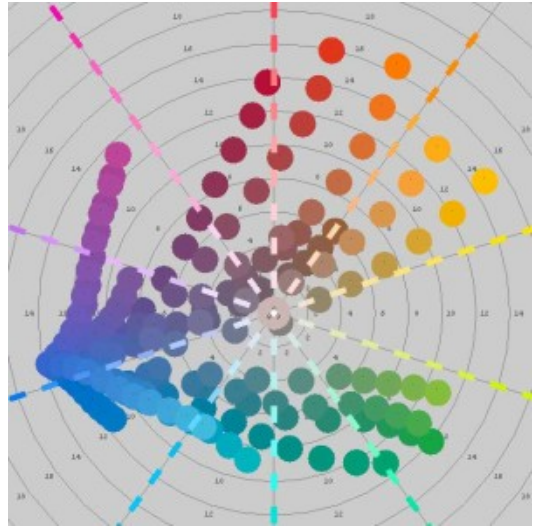

| Green (#312 Light Green Permanent) mixing with others. | Blue (#381 Cobalt Blue Hue) mixing with others. |

It appears that the WGM mixing rule produces very reasonable results! Furthermore, I was pleased to see that some of the "octopus" shapes that David Briggs described are exhibited by the WGM mix, although to a somewhat lesser extent than he observed. I'm not sure what mixing rule is used in rs2color; I'm assuming it is based on a much more advanced theory of paint mixture (such as the Kubelka-Munk theory[13-14]). In comparison to the rs2color results, it appears that the WGM rule is a promising simplified procedure.

Incidentally, mixing with black and white also works well. Here are mixing paths for each of the Liquitex colors mixing with:

o   Titanium White

   (reflectance = 0.1228, 0.2032, 0.3886, 0.6489, 0.8518, 0.9362, 0.9568, 0.9625, 0.9673, 0.9678, 0.9677, 0.9694, 0.9691, 0.9691, 0.9701, 0.9692, 0.9692, 0.9693, 0.9668, 0.9695, 0.9679, 0.9676, 0.9671, 0.9673, 0.96734, 0.9655, 0.9661, 0.9676, 0.9700, 0.9694, 0.9680, 0.9678, 0.9692, 0.9704, 0.9705, 0.9730)

o   Ivory Black

   (reflectance = 0.0298, 0.0466, 0.0635, 0.0803, 0.0931, 0.0957, 0.0984, 0.1028, 0.1077, 0.1129, 0.1183, 0.1208, 0.1210, 0.1225, 0.1251, 0.1274, 0.1300, 0.1325, 0.1347, 0.1374, 0.1394, 0.1421, 0.1442, 0.1456, 0.1472, 0.1493, 0.1517, 0.1537, 0.1561, 0.1579, 0.1602, 0.1622, 0.1642, 0.1669, 0.1690, 0.1711)

using intermediate mix ratios of 1:9, 2:8, 3:7, ... , 9:1

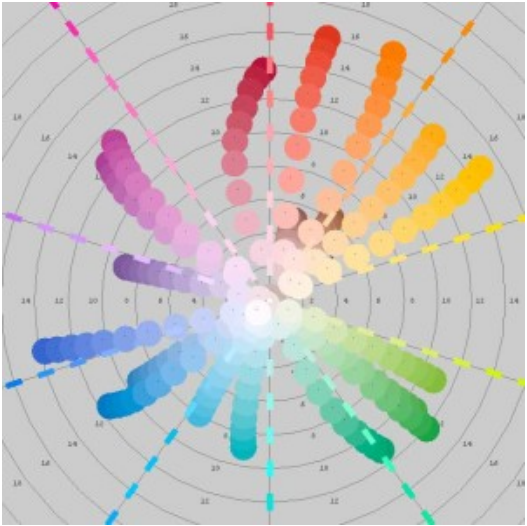 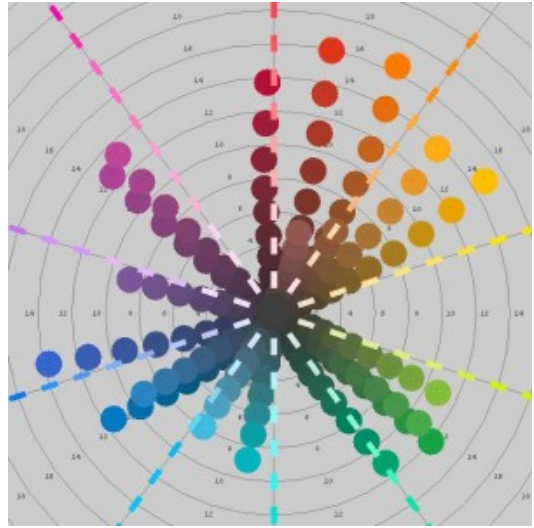

Tints of all colors with titanium white.   Shades of all colors with ivory black.

Be careful not to use zeros for the reflectance curve for black. That represents a colorant with an infinite amount of shading power, which turns the mix black regardless of how little is used!

At this point, you might be wondering, "That's great, but what about mixing RGB colors instead of reflectance curves?" That is indeed our ultimate goal. I see several approaches we can take to that end.

First, we can look up reflectance curves from RGB values using catalogs of published reflectance curves. This is the easiest approach, but has some pitfalls. We are forced to choose catalog colors that are nearest to our source colors, which may not be near enough for our purposes. Also, this is not a practical solution if we are mixing a large number of colors or if we wish to create a smooth gradient of mixed colors.

The second alternative is to compute a reflectance curve directly from the source sRGB values. This is more complex but appears to be do-able. This is what I consider to be my "original contribution" to this area of study.

Let's consider the two cases separately in the following sections.

**Cataloged Reflectance Curves**

One of the most widely available catalogs of reflectance curves comes from the [Munsell Color System](#) publications.[15] This system describes object colors by hue, chroma (similar to saturation), and value (similar to brightness).

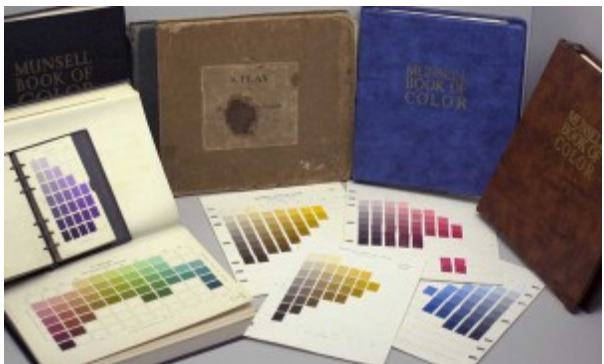
Various editions of the Munsell Book of Colors
(By Mark Fairchild [CC BY-SA 3.0] via Wikimedia Commons)

Paul Centore published this web site, which has links to various catalogs of measured reflectance curves of the Munsell samples.[16] In particular, I've found the data in this text file[17], which lists 1485 different Munsell colors, to be most useful in my studies. I'm assuming these measurements were made by Paul Centore. I've created an Excel spreadsheet[18] for this data that may be more convenient than the text file version.

I've also added some columns to the spreadsheet containing sRGB values for each of the Munsell colors when viewed under D65 illumination, using the equations presented earlier. (Here is a text file of the sRGB values in case the Excel format is not handy. They are presented in the same order as the reflection curves in the previous text file and Excel file.) You might notice that some of the sRGB values all outside the range 0 to 255. That is because these colors fall outside the sRGB gamut. This should not cause a problem since we are using good sRGB values to find a reflectance curve from this data, not sRGB values *from* a reflectance curve.

Suppose we have two sRGB colors we wish to mix subtractively. We can do this by finding the nearest Munsell color to each of these sRGB colors, and then mixing the corresponding reflectance curves. A little software is handy here. An exhaustive search could be performed by computing the distance between the given sRGB and each of the Munsell sRGB triplets, that is,

$$dist^2 = (R_{given} - R_{Munsell})^2 + (G_{given} - G_{Munsell})^2 + (B_{given} - B_{Munsell})^2,$$

and then selecting the smallest distance. Or, in the Excel spreadsheet, a new column could be defined that computes this sum of squares, and then the rows could be sorted by this column to bring the smallest to the top. There are other more efficient algorithms for doing nearest-neighbor searches. In Matlab, I would generate a Delaunay triangulation of the sRGB Munsell values and then use the nearest neighbor function to operate on it.

The quality of the selection can be improved somewhat if we use a different color space. sRGB is not considered "perceptually uniform," that is, equal movements in sRGB space do not represent equal changes in perceived color. In some parts of the space, large changes in color happen with small changes in sRGB values, and vice versa. The L*a*b* color space is more perceptually uniform. So it would be better to convert the sRGB val-ues to L*a*b* values (by doing this, then this), and then compute the distances between the given L*a*b* triplet and the Munsell L*a*b* values.

It should be noted that the Munsell samples are intended to be viewed under Standard Illuminant C instead of D65. Illuminant C is intended to mimic northern sky daylight, whereas D65 mimics noon daylight. They are not far apart in color temperature: C is 6770 K and D65 is 6500 K. However, for our purposes, the intended viewing conditions are irrelevant. We are simply using the Munsell dataset as a source for reflectance curves corresponding to specified D65-referenced sRGB values. The hue/chroma/value designations of the Munsell samples are irrelevant. The Munsell set is simply providing us with a pool of reflectance curves that belong to real painted surfaces.

Once the Munsell reflectance curves most closely matching the sRGB colors being mixed are selected, then the reflectance curves can be mixed using the weighted geometric mean mixing rule presented earlier. The resulting reflectance curve can then be converted to sRGB to complete the mixing process.

There is a possibility that the mixed curve will give an sRGB value outside the sRGB gamut (outside 0-255). In that case, I'd advise simply clipping the values to stay within 0 and 255. However, I have a gut feeling that if the two colors being mixed are within the sRGB gamut, then the weighted geometric mean mixture will also be within gamut. This might be easy to prove, but I'm not going to take the time to investigate right now.

**Computing Reflectance Curves Directly from sRGB Values**

Instead of relying on existing reflectance measurement data, it is possible to generate reflectance curves directly from sRGB values, and then use these curves in WGM mixing calculations. The main difficulty, however, is that there are an infinite number of different reflectance curves that all give rise to the same color sensation, i.e., the same sRGB color. From a math standpoint, this is evident in the shape of the $T$ matrix; it has many more columns than rows, making the linear system underdetermined.

While it is not too difficult to find a single reflectance curve with a specific sRGB value, it may not be suitable for subtractive color mixture computations. For example, a reflectance curve comprising a handful of spiked values at various wavelengths would give awful color mixture results. Or, a reflectance curve with negative values, while mathematically giving the correct sRGB value, would cause WGM calculations to fail completely (raising a negative number to a fractional power is prohibited in real-valued calculations).

I've recently developed a set of algorithms that compute reflectance curves from sRGB triplets that give good quality results. By "good quality" I mean they produce reflectance curves quite similar to those of colored objects found in nature, specifically those associated with commercial artist's paints or color pigments used in those paints. More information can be found at here.[19] Five algorithms are presented there, three of which I recommend for subtractive color mixture computations. Here is a comparison of the three:

| Algorithm Name | Computational Effort | Comments | Link to Matlab/ Octave Code |
|---|---|---|---|
| **ILSS (Iterative Least Slope Squared)** | Relatively little. | Very fast, but tends to undershoot reflectance curve peaks, especially for bright red and purple colors. Always returns reflectance values in the range 0-1. | link |
| **LLSS (Least Log Slope Squared)** | About 12 times that of ILSS. | Better quality matches overall, but tends to overshoot peaks in the yellow region. Some reflectance values can be >1, especially for bright red colors. | link |
| **ILLSS (Iterative Least Log Slope Squared)** | About 20 times that of ILSS. | Best quality matches. Tends to overshoot peaks in the yellow region. Always returns reflectance values in the range 0-1. | link |

The code for each of the algorithms runs in either Matlab, or the free open source alternative called Octave.

Here are the six Liquitex colors investigated earlier, showing the original measured reflectance (blue curve) and the reflectance generated by the LLSS algorithm (red curve):

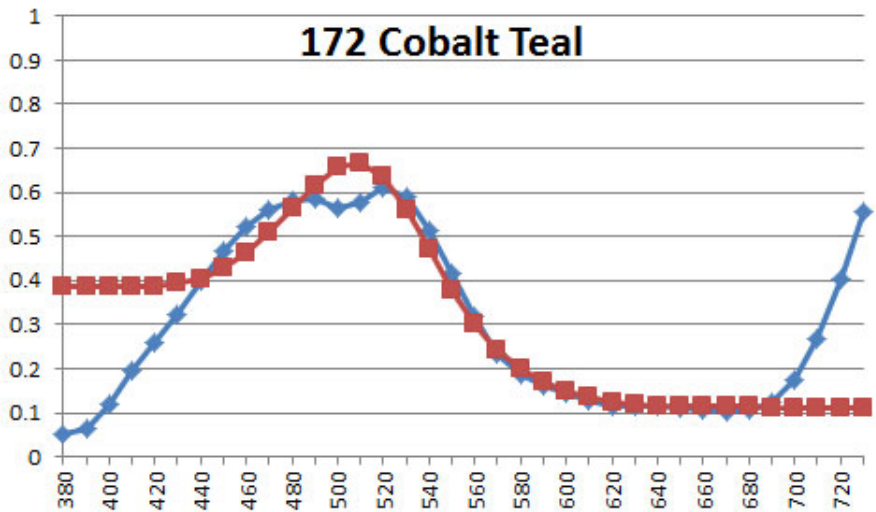

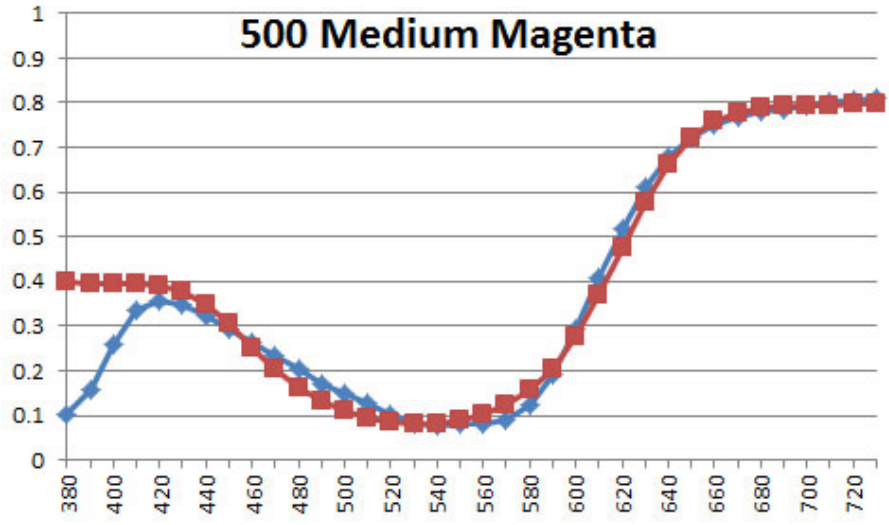

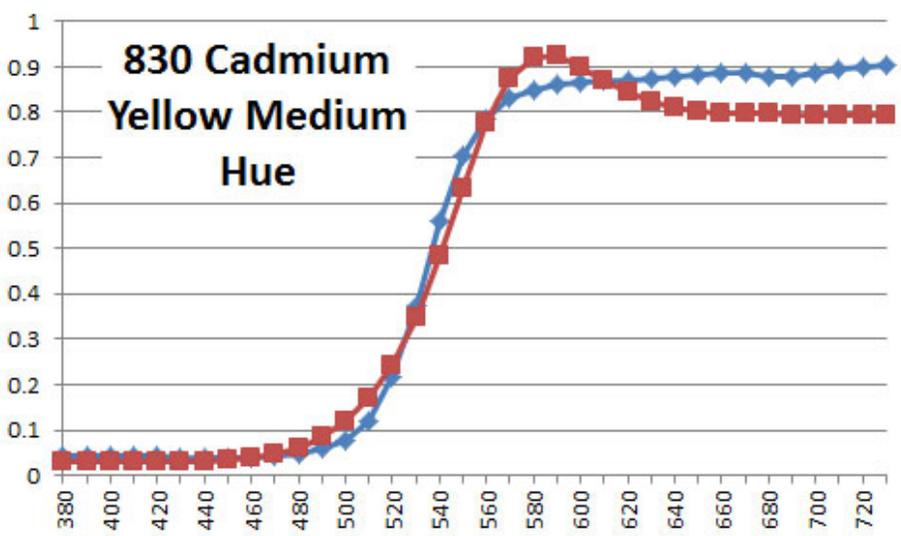

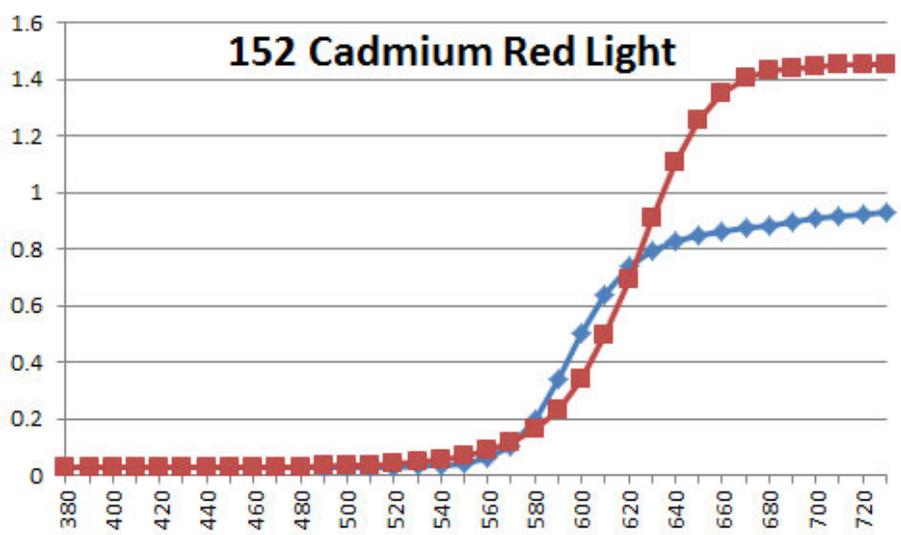

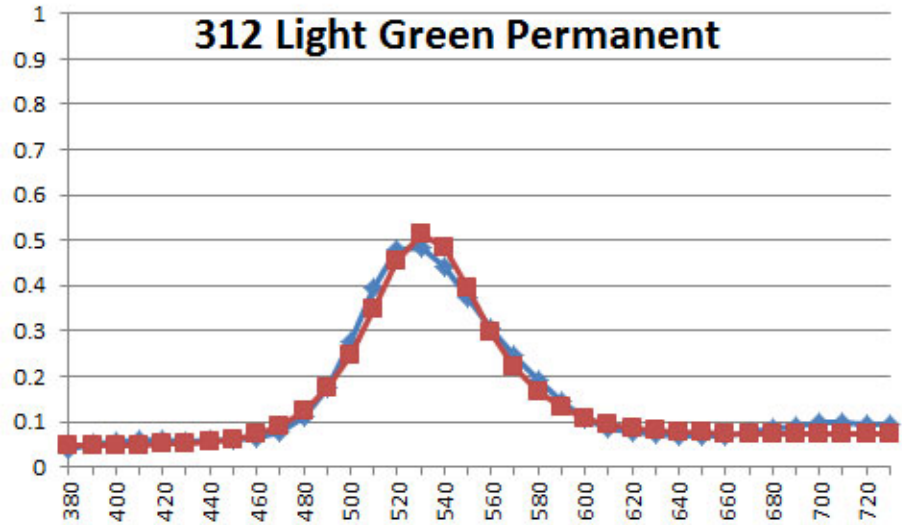

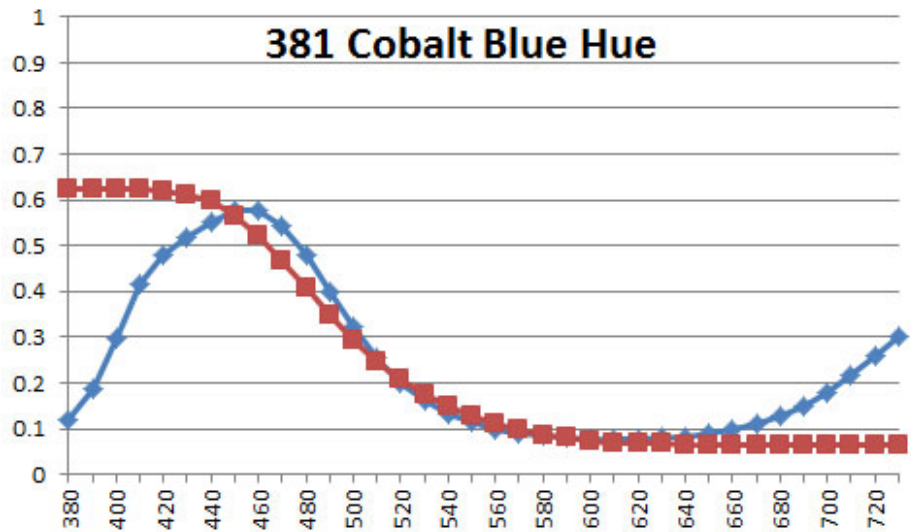

Notice how there tends to be more discrepancy between the original reflectance curve and the generated one at the very high and low wavelengths. Human vision is far less sensitive to these outer wavelengths, so these discrepancies have little impact on perceived color. Keep in mind that even though the two curves may differ considerably at the ends, they both give the identical sRGB values and perceived color.

Your choice of which algorithm to use depends on your specific needs. If computational efficiency is more important than realistic color mixing, use ILSS. For best results at the expense of much more computation, use ILLSS. The LLSS method offers a balance of good results and moderately high computational effort.

Another factor that may influence the decision of which algorithm to use is one of aesthetics. It is a common expectation that mixing blue and yellow subtractively will give some type of green, instead of the neutral gray that comes from additive color mixture. The difference in the behavior of the various methods has an impact on what kind of green is produced. The figure shows an example of mixing yellow (255, 255, 0) and blue (0, 0, 255) in various proportions.

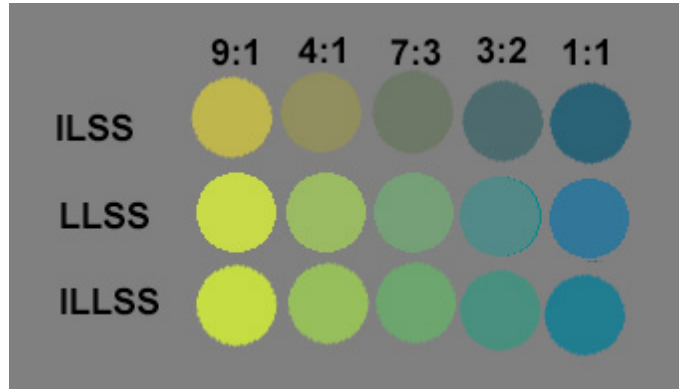

A comparison of the type of greens that are obtained from mixing yellow and blue in varying proportions by ILSS, LLSS, and ILLSS.

It appears that ILLSS and LLSS give brighter and more chromatic greens in comparison to ILSS, which may sway favor toward those methods if the high computational requirements can be tolerated. (Incidentally, if you're looking for a more powerful green in subtractive mixture, try yellow and cyan instead of yellow and blue!)

**Other Applications**

The work presented here is also applicable to other color spaces, other color matching functions (observers), and other reference illuminants. Recall we started with

$$\{rgb\} = [M][A']\{N\}/w.$$

To accommodate other color spaces, the $M$ matrix would be changed. For example, to operate in $XYZ$ space, simply use the identity matrix for $M$. For different RGB color spaces, create a new $M$ matrix from the RGB primaries and reference white according to [this calculation](#).[20] Different standard observers would require different $A'$ matrices, such as the CIE 1964 10 degree color matching functions. [Here is a site](#)[21] compiling many different color matching function sets. Finally, different illuminants would be treated in the computation of $N$, which is $\{N\} = [W]\{\rho\}$. A different illuminant would be implemented by placing it along the diagonal of the $W$ matrix. [The Munsell Color Science Laboratory](#)[22] has links to many different standard illuminants (and a wealth of other data).

___________________________________________

**Appendix: Linked Textual Data**

Several data tables and source codes are supplied in the text above via internet links. For archival purposes, these tables and codes are supplied below.

A' matrix (3x36)

```
CIE 1931 color matching functions for 380 to 730 nm by 10 nm intervals

0.001368, 0.004243, 0.01431, 0.04351, 0.13438, 0.2839, 0.34828, 0.3362,
0.2908, 0.19536, 0.09564, 0.03201, 0.0049, 0.0093, 0.06327, 0.1655, 0.2904,
0.43345, 0.5945, 0.7621, 0.9163, 1.0263, 1.0622, 1.0026, 0.85445, 0.6424,
0.4479, 0.2835, 0.1649, 0.0874, 0.04677, 0.0227, 0.011359, 0.00579, 0.002899,
0.00144

0.000039, 0.00012, 0.000396, 0.00121, 0.004, 0.0116, 0.023, 0.038, 0.06,
0.09098, 0.13902, 0.20802, 0.323, 0.503, 0.71, 0.862, 0.954, 0.99495, 0.995,
0.952, 0.87, 0.757, 0.631, 0.503, 0.381, 0.265, 0.175, 0.107, 0.061, 0.032,
0.017, 0.00821, 0.004102, 0.002091, 0.001047, 0.00052
```

```
0.00645, 0.02005, 0.06785, 0.2074, 0.6456, 1.3856, 1.74706, 1.77211, 1.6692,
1.28764, 0.81295, 0.46518, 0.272, 0.1582, 0.07825, 0.04216, 0.0203, 0.00875,
0.0039, 0.0021, 0.00165, 0.0011, 0.0008, 0.00034, 0.00019, 0.00005, 0.00002,
0, 0, 0, 0, 0, 0, 0, 0, 0
```

M matrix (3x3)

Conversion between tristimulus values, XYZ, and linear rgb, referenced to D65 illuminant

```
3.243063328, -1.538376194, -0.49893282
-0.968963091, 1.875424508, 0.041543029
0.055683923, -0.204174384, 1.057994536
```

T matrix (3x36)

```
5.47813E-05, 0.000184722, 0.000935514, 0.003096265, 0.009507714, 0.017351596,
0.022073595, 0.016353161, 0.002002407, -0.016177731, -0.033929391, -
0.046158952, -0.06381706, -0.083911194, -0.091832385, -0.08258148, -
0.052950086, -0.012727224, 0.037413037, 0.091701812, 0.147964686,
0.181542886, 0.210684154, 0.210058081, 0.181312094, 0.132064724, 0.093723787,
0.057159281, 0.033469657, 0.018235464, 0.009298756, 0.004023687, 0.002068643,
0.00109484, 0.000454231, 0.000255925

-4.65552E-05, -0.000157894, -0.000806935, -0.002707449, -0.008477628, -
0.016058258, -0.02200529, -0.020027434, -0.011137726, 0.003784809,
0.022138944, 0.038965605, 0.063361718, 0.095981626, 0.126280277, 0.148575844,
0.149044804, 0.14239936, 0.122084916, 0.09544734, 0.067421931, 0.035691251,
0.01313278, -0.002384996, -0.009409573, -0.009888983, -0.008379513, -
0.005606153, -0.003444663, -0.001921041, -0.000995333, -0.000435322, -
0.000224537, -0.000118838, -4.93038E-05, -2.77789E-05

0.00032594, 0.001107914, 0.005677477, 0.01918448, 0.060978641, 0.121348231,
0.184875618, 0.208804428, 0.197318551, 0.147233899, 0.091819086, 0.046485543,
0.022982618, 0.00665036, -0.005816014, -0.012450334, -0.015524259, -
0.016712927, -0.01570093, -0.013647887, -0.011317812, -0.008077223, -
0.005863171, -0.003943485, -0.002490472, -0.001440876, -0.000852895, -
0.000458929, -0.000248389, -0.000129773, -6.41985E-05, -2.71982E-05, -
1.38913E-05, -7.35203E-06, -3.05024E-06, -1.71858E-06
```

ILSS (Iterative Least Slope Squared) source code (Matlab and Octave)

```
function rho=ILSS(B11,B12,sRGB)
% This is the Iterative Least Slope Squared (ILSS) algorithm for generating
% a "reasonable" reflectance curve from a given sRGB color triplet.
% The reflectance spans the wavelength range 380-730 nm in 10 nm increments.

% It solves
% min   sum(rho_i+1 - rho_i)^2
% s.t. T rho = rgb,
%      K1 rho = 1,
%      K0 rho = 0,
% using Lagrangian formulation and iteration to keep all rho (0-1].

% B11 is upper-left 36x36 part of inv([D,T';T,zeros(3)])
```

```
% B12 is upper-right 36x3 part of inv([D,T';T,zeros(3)])
% sRGB is a 3-element vector of target D65-referenced sRGB values (0-255),
% rho is a 36x1 vector of reflectance values (0->1) over
%     wavelengths 380-730 nm,

% Written by Scott Allen Burns, 4/26/15.
% Licensed under a Creative Commons Attribution-ShareAlike 4.0 International
% License (http://creativecommons.org/licenses/by-sa/4.0/).
% For more information, see
% http://www.scottburns.us/subtractive-color-mixture/

rho=ones(36,1)/2; % initialize output to 0.5
rhomin=0.00001; % smallest refl value

% handle special case of (255,255,255)
if all(sRGB==255)
    rho=ones(36,1);
    return
end

% handle special case of (0,0,0)
if all(sRGB==0)
    rho=rhomin*ones(36,1);
    return
end

% compute target linear rgb values
sRGB=sRGB(:)/255; % convert to 0-1 column vector
rgb=zeros(3,1);
% remove gamma correction to get linear rgb
for i=1:3
    if sRGB(i)<0.04045
        rgb(i)=sRGB(i)/12.92;
    else
        rgb(i)=((sRGB(i)+0.055)/1.055)^2.4;
    end
end

R=B12*rgb;

% iteration to get all refl 0-1
maxit=10; % max iterations
count=0; % counter for iteration
while ( (any(rho>1) || any(rho<rhomin)) && count<=maxit ) || count==0
    % create K1 matrix for fixed refl at 1
    fixed_upper_logical = rho>=1;
    fixed_upper=find(fixed_upper_logical);
    num_upper=length(fixed_upper);
    K1=zeros(num_upper,36);
    for i=1:num_upper
        K1(i,fixed_upper(i))=1;
    end

    % create K0 matrix for fixed refl at rhomin
    fixed_lower_logical = rho<=rhomin;
    fixed_lower=find(fixed_lower_logical);
    num_lower=length(fixed_lower);
```

```
    K0=zeros(num_lower,36);
    for i=1:num_lower
        K0(i,fixed_lower(i))=1;
    end

    % set up linear system
    K=[K1;K0];
    C=B11*K'/(K*B11*K'); % M*K'*inv(K*M*K')
    rho=R-C*(K*R-[ones(num_upper,1);rhomin*ones(num_lower,1)]);
    rho(fixed_upper_logical)=1; % eliminate FP noise
    rho(fixed_lower_logical)=rhomin; % eliminate FP noise

    count=count+1;
end
if count>=maxit
    disp(['No solution found after ',num2str(maxit),' iterations.'])
end
```

LLSS (Least Log Slope Squared) source code (Matlab and Octave)

```
function rho=LLSS(T,sRGB)
% This is the Least Log Slope Squared (LLSS) algorithm for generating
% a "reasonable" reflectance curve from a given sRGB color triplet.
% The reflectance spans the wavelength range 380-730 nm in 10 nm increments.

% Solves min sum(z_i+1 - z_i)^2 s.t. T exp(z) = rgb, where
% z=log(reflectance), using Lagrangian formulation and Newton's method.
% Allows reflectance values >1 to be in solution.

% T is 3x36 matrix converting reflectance to D65-weighted linear rgb,
% sRGB is a 3 element vector of target D65 referenced sRGB values (0-255),
% rho is a 36x1 vector of reconstructed reflectance values, all > 0,

% For more information, see
% http://www.scottburns.us/subtractive-color-mixture/
% Written by Scott Allen Burns, March 2015.
% Licensed under a Creative Commons Attribution-ShareAlike 4.0 International
% License (http://creativecommons.org/licenses/by-sa/4.0/).

% initialize outputs to zeros
rho=zeros(36,1);

% handle special case of (0,0,0)
if all(sRGB==0)
    rho=0.0001*ones(36,1);
    return
end

% 36x36 difference matrix for Jacobian
% having 4 on main diagonal and -2 on off diagonals,
% except first and last main diagonal are 2.
D=full(gallery('tridiag',36,-2,4,-2));
D(1,1)=2;
D(36,36)=2;

% compute target linear rgb values
```

```
sRGB=sRGB(:)/255; % convert to 0-1
rgb=zeros(3,1);
% remove gamma correction to get linear rgb
for i=1:3
    if sRGB(i)<0.04045
        rgb(i)=sRGB(i)/12.92;
    else
        rgb(i)=((sRGB(i)+0.055)/1.055)^2.4;
    end
end

% initialize
z=zeros(36,1); % starting point all zeros
lambda=zeros(3,1); % starting Lagrange mult
maxit=100; % max number of iterations
ftol=1.0e-8; % function solution tolerance
deltatol=1.0e-8; % change in oper pt tolerance
count=0; % iteration counter

% Newton's method iteration
while count <= maxit
    r=exp(z);
    v=-diag(r)*T'*lambda; % 36x1
    m1=-T*r; % 3x1
    m2=-T*diag(r); % 3x36
    F=[D*z+v;m1+rgb]; % 39x1 function vector
    J=[D+diag(v),m2';m2,zeros(3)]; % 39x39 Jacobian matrix
    delta=J\(-F); % solve Newton system of equations J*delta = -F
    z=z+delta(1:36); % update z
    lambda=lambda+delta(37:39); % update lambda
    if all(abs(F)<ftol) % check if functions satisfied
        if all(abs(delta)<deltatol) % check if variables converged
            % solution found
            disp(['Solution found after ',num2str(count),' iterations'])
            rho=exp(z);
            return
        end
    end
    count=count+1;
end
disp(['No solution found in ',num2str(maxit),' iterations.'])
```

ILLSS (Iterative Least Log Slope Squared) source code (Matlab and Octave)

```
function rho=ILLSS(T,sRGB)
% This is the Iterative Least Log Slope Squared (ILLSS) algorithm for
% generating a "reasonable" reflectance curve from a given sRGB color
% triplet. The reflectance spans the wavelength range 380-730 nm in 10 nm
% increments.

% It solves min sum(z_i+1 - z_i)^2 s.t. T exp(z) = rgb, K z = 0, where
% z=log(reflectance), using Lagrangian approach and Newton's method.
% Clips values >1 and repeats optimization until all reflectance <=1.

% T is 3x36 matrix converting reflectance to linear rgb over the
%   range 380-730 nm,
```

```
% sRGB is a 3 element vector of target D65 referenced sRGB values
%       in 0-255 range,
% rho is a 36x1 vector of reflectance values (0->1] over
%       wavelengths 380-730 nm,

% Written by Scott Allen Burns, 4/11/15.
% Licensed under a Creative Commons Attribution-ShareAlike 4.0 International
% License (http://creativecommons.org/licenses/by-sa/4.0/).
% For more information, see
% http://www.scottburns.us/subtractive-color-mixture/

% initialize output to zeros
rho=zeros(36,1);

% handle special case of (0,0,0)
if all(sRGB==0)
    rho=0.0001*ones(36,1);
    return
end

% handle special case of (255,255,255)
if all(sRGB==255)
    rho=ones(36,1);
    return
end

% 36x36 difference matrix having 4 on main diagonal and -2 on off diagonals,
% except first and last main diagonal are 2.
D=full(gallery('tridiag',36,-2,4,-2));
D(1,1)=2;
D(36,36)=2;

% compute target linear rgb values
sRGB=sRGB(:)/255; % convert to 0-1 column vector
rgb=zeros(3,1);
% remove gamma correction to get linear rgb
for i=1:3
    if sRGB(i)<0.04045
        rgb(i)=sRGB(i)/12.92;
    else
        rgb(i)=((sRGB(i)+0.055)/1.055)^2.4;
    end
end

% outer iteration to get all refl <=1
maxouter=10;
outer_count=0; % counter for outer iteration
while (any(rho>1) && outer_count<=maxouter) || all(rho==0)
    % create K matrix for fixed refl constraints
    fixed_refl=find(rho>=1)';
    numfixed=length(fixed_refl);
    K=zeros(numfixed,36);
    for i=1:numfixed
        K(i,fixed_refl(i))=1;
    end

    % initialize
```

```matlab
        z=zeros(36,1); % starting point all zeros
        lambda=zeros(3,1); % starting point for lambda
        mu=zeros(numfixed,1); % starting point for mu
        maxit=50; % max number of iterations
        ftol=1.0e-8; % function solution tolerance
        deltatol=1.0e-8; % change in oper pt tolerance
        count=0; % iteration counter

        % Newton's method iteration
        while count <= maxit
            r=exp(z);
            v=-diag(r)*T'*lambda; % 36x1
            m1=-T*r; % 3x1
            m2=-T*diag(r); % 3x36
            F=[D*z+v+K'*mu;m1+rgb;K*z]; % function vector
            J=[D+diag(v),[m2',K'];[m2;K],zeros(numfixed+3)]; % Jacobian matrix
            delta=J\(-F); % solve Newton system of equations J*delta = -F
            z=z+delta(1:36); % update z
            lambda=lambda+delta(37:39); % update lambda
            mu=mu+delta(40:end);
            if all(abs(F)<ftol) % check if functions satisfied
                if all(abs(delta)<deltatol) % check if variables converged
                    % solution found
                    disp(['Inner loop solution found after ',num2str(count),...
                        ' iterations'])
                    rho=exp(z);
                    break
                end
            end
            count=count+1;
        end
        if count>=maxit
            disp(['No inner loop solution found after ',num2str(maxit),...
                ' iterations.'])
        end
        outer_count=outer_count+1;
end
if outer_count<maxouter
    disp(['Outer loop solution found after ',num2str(outer_count),...
        ' iterations'])
else
    disp(['No outer loop solution found after ',num2str(maxouter),...
        ' iterations.'])
end
```